\documentclass[letterpaper,12pt]{article}
\usepackage{tabularx} 
\usepackage{amsmath} 
\usepackage{graphicx}
\usepackage[margin=1in,letterpaper]{geometry}
\usepackage{tikz}
\usepackage[italic]{hepnames}
\usepackage{lscape}
\usepackage{epsfig}
\usepackage{float}
\usepackage{booktabs}
\usepackage{authblk}
\usepackage{color}
\usepackage[percent]{overpic}
\usepackage{subfig}
\usepackage{cite} 
\usepackage[final]{hyperref} 
\hypersetup{
    colorlinks=true,      
    linkcolor=blue,       
    citecolor=red,        
    filecolor=magenta,     
    urlcolor=pink         
}

\begin{document}
\title{Degeneracy resolution capabilities of NO$\nu$A and DUNE in the presence of light sterile neutrino}

\author[a]{\normalsize{Akshay Chatla}\thanks{email: chatlaakshay@gmail.com}}
\author[b]{Sahithi Rudrabhatla\thanks{email: srudra4@uic.edu}}
\author[a]{{Bindu A. Bambah}\thanks{email: bbambah@gmail.com}}

\affil[a]{\small{School of Physics, University of Hyderabad, Hyderabad - $500046$, India}}
\affil[b]{Department of Physics, University of Illinois at Chicago, Chicago, IL, $60607$}
\date{\today}
\maketitle
\begin{abstract}
We investigate the implications of a sterile neutrino on the physics potential of the proposed experiment DUNE and future runs of NO$\nu$A using latest NO$\nu$A results. Using combined analysis of the disappearance and appearance data, NO$\nu$A reported preferred solutions at normal hierarchy (NH)  with two degenerate best-fit points one in the lower octant (LO) and $\delta_{13}$ = 1.48$\pi$ and other in higher octant (HO) and $\delta_{13}$ = 0.74$\pi$. Another solution of inverted hierarchy (IH) which is 0.46$\sigma$ away from best fit was also reported. We discuss chances of resolving these degeneracies in the presence of sterile neutrino.

\end{abstract}

\section{Introduction}
Sterile neutrinos are hypothetical particles that do not interact via any of the fundamental interactions other than gravity. The term sterile is used to distinguish them from active neutrinos, which are charged under weak interaction. The theoretical motivation for sterile neutrino explains the active neutrino
mass after spontaneous symmetry breaking, by adding a gauge singlet term
(sterile neutrino) to the Lagrangian under $SU(3)_c \otimes SU(2)_L \otimes U(1)_r$ where the Dirac term appears through the Higgs mechanism, and Majorana mass term is a
gauge singlet, and hence appears as a bare
mass term\cite{Volkas:2001zb}. The diagonalization of the mass matrix gives masses to all neutrinos due to the See-Saw mechanism.

Some experimental anomalies also point towards the existence of sterile neutrinos.
Liquid Scintillator Neutrino Detector(LSND) detected
$\overline{\nu}_{\mu} \rightarrow\overline{\nu}_{e}$ 
 transitions indicating $\Delta m^2 \approx 1eV^2$
which is inconsistent with $\Delta m^2_{32},\Delta m^2_{21}$ (LSND anomaly)\cite{Athanassopoulos:1995iw}. Measurement of the width of Z boson by LEP gave number of active neutrinos to be 2.984$\pm$0.008\cite{ALEPH:2005ab}. Thus the new neutrino introduced to explain the anomaly has to be a sterile neutrino. MiniBooNE, designed to verify the LSND anomaly, observed an unexplained excess of events in low-energy region of  $\overline{\nu}_{e},{\nu}_{e}$ spectra, consistent with LSND \cite{AguilarArevalo:2008rc}. SAGE and GALLEX observed  lower event rate than expected,  explained by the oscillations of $\nu_{e}$ due to $\Delta m^2 \geq 1eV^2$(Gallium anomaly)\cite{Abdurashitov:2005tb,Acero:2007su,Giunti:2010zu}. Recent precise predictions of reactor anti-neutrino flux has increased the expected flux by 3$\%$ over old predictions. With the new flux evaluation, the ratio of observed and predicted flux deviates at 98.6 $\%$ C.L(Confidence level) from unity, this is called ``Reactor anti-neutrino Anomaly"\cite{Mention:2011rk}. This anomaly can also be explained using  sterile neutrino model.

\medskip
Short-baseline(SBL) experiments are running to search for sterile neutrinos. SBL experiments are the best place to look for sterile neutrino, as they are sensitive to new expected mass-squared splitting $\Delta \rm{m}^{2} \simeq 1$eV$^2$. However, SBL experiments cannot study all the properties of sterile neutrinos, mainly new CP phases introduced by sterile neutrino models. These new CP phases need long distances to become measurable\cite{Klop:2014ima, Palazzo:2015gja}, thus can be measured using Long baseline(LBL) experiments. With the discovery of relatively large value for $\theta_{13}$  by Daya Bay\cite{An:2014ehw}, the sensitivity of LBL experiments towards neutrino mass hierarchy and CP phases increased significantly. In this  context,  some  phenomenological  studies  regarding  the  sensitivity  of
LBL experiments can be found in recent works\cite{Soumya:2016aif,C.:2014ika,Deepthi:2014iya,Ghosh:2015ena,Goswami:2017hcw}. Using recent global fits of oscillation parameters in the 3+1 scenario\cite{Kopp:2013vaa}, current LBL experiments can extract two out of three CP phases (one of them being standard $\delta_{13}$)\cite{Palazzo:2015gja}. The phenomenological  studies of LBL experiments in presence of sterile neutrino is studied by several groups\cite{Ghosh:2017atj,Agarwalla:2016xxa,Dutta:2016glq,Agarwalla:2018,Agarwalla:2016xlg,Choubey:2017ppj}. Now, the sensitivity of LBL experiments towards their original goals decreases due to sterile neutrinos. It is seen in case of the CPV measurement; new CP-phases will decrease the sensitivity towards standard CP phase ($\delta_{13}$). This will reduce degeneracy resolution capacities of LBL experiments. In this paper, we study hierarchy-$\theta_{23}$-$\delta_{13}$ degeneracies using contours in $\theta_{23}$-$\delta_{13}$ plane and how they are affected by the introduction of sterile neutrinos. We attempt to find the extent to which these degeneracies can be resolved in future runs of NO$\nu$A and DUNE.

\medskip 
The outline of the paper is as follows. In section 2, we present the experimental specifications of NO$\nu$A and DUNE used in our simulation. We introduce the effect of sterile neutrino on parameter degeneracies resolution in section 3. Section 4 contains the discussion about the degeneracy resolving capacities of future runs of NO$\nu$A and DUNE assuming latest NO$\nu$A results - NH(Normal hierarchy)-LO(Lower octant), NH-HO(Higher octant), and IH(Inverted hierarchy)-HO as true solutions for both 3 and 3+1 models. Finally, Section 5 contains concluding comments on our results.

\section{Experiment specifications}
We used GLoBES (General Long Baseline Experiment simulator) \cite{Huber:2004ka, Huber:2007ji} to simulate the data for different LBL experiments including NO$\nu$A and DUNE. The neutrino oscillation probabilities for the 3+1 model are calculated using the new physics engine available from Ref.\cite{globes3}.

NO$\nu$A\cite{Adamson:2016xxw, Adamson:2016tbq} is an LBL experiment which started its full operation from October 2014. NO$\nu$A has two detectors, the near detector is located at Fermilab (300 ton, 1 km from NuMI beam target) while the far detector(14 Kt) is located at Northern Minnesota 14.6 mrad off the NuMI beam axis at 810 km from NuMI beam target, justifying ``Off-Axis" in the name. This off-axis orientation gives us a narrow beam of flux, peak at 2 GeV\cite{Ayres:2007tu}. For simulations, we used NO$\nu$A setup from Ref.\cite{Agarwalla:2012bv}. We used the full projected exposure of 3.6 x $ 10^{21}$ p.o.t (protons on target) expected after six years of runtime at 700kW beam power. Assuming the same runtime for neutrino and anti-neutrino modes, we get 1.8 x $ 10^{21}$ p.o.t for each mode. Following \cite{Ayres:2004js} we considered 5$\%$ normalization error for the signal, 10 $\%$ error for the background for appearance and disappearance channels.

\medskip                                                                                                                    
DUNE (Deep Underground Neutrino Experiment)\cite{Acciarri:2015uup,Acciarri:2016crz} is the next generation LBL experiment. Long Base Neutrino Facility(LBNF) of Fermilab is the source for DUNE. Near detector of DUNE will be at Fermilab. Liquid Argon detector of 40 kt to be constructed at Sanford Underground Research Facility situated 1300 km from the beam target, will act as the far detector. DUNE uses the same source as of NO$\nu$A; we will observe beam flux peak at 2.5GeV. We used DUNE setup give in Ref.\cite{Alion:2016uaj} for our simulations. Since DUNE is still in its early stages, we used simplified systematic treatment, i.e., 5$\%$ normalization error on signal, 10 $\%$ error on the background for both appearance and disappearance spectra. We give experimental details described above in tabular form in tables \ref{tab:table1} and \ref{tab:table2}. 
\medskip

Oscillation parameters are estimated from the data by comparing observed and predicted $\nu_{e}$ and $\nu_{\mu}$ interaction rates and energy spectra. 
GLoBES calculates event rates of neutrinos for energy bins taking systematic errors, detector resolutions, MSW effect due to earth's crust etc into account. The event rates generated for true and test values are used to plot   $\chi^2$ contours. GLoBES uses its inbuilt algorithm to calculate $\chi^2$ values numerically considering parameter correlations as well as systematic errors. In our calculations we used $\chi^2$ as:

\begin{equation}
\chi^2 = \rm{\sum_{i=1}^{\# of bins}\sum_{E_n=E_{1},E_{2}..} \frac{(O_{E_n,i} - (1+a_{F} +a_{E_n})T_{E,i})^2}{O_{E_n,i}} + \frac{a_{F}^2}{\sigma_{F}^2} + \frac{a_{E_n}^2}{\sigma_{E_n}^2}}
\end{equation}
where $\rm{O}_{\rm{E}_1,\rm{i}},\rm{O}_{\rm{E}_2,\rm{i}}..$  are the event rates for the $\rm{i}^{\rm{th}}$ bin in the detectors of different experiments, calculated for true values of oscillation parameters; $\rm{T_{E_n,i}}$ are the expected event rates for the $i^{th}$ bin in the detectors of different experiments for the test parameter values; $\rm{a_F , a_{E_n}}$ are the uncertainties associated
with the flux and detector mass; and $\sigma_{F},\sigma_{E_n}$ are the respective associated standard
deviations. The calculated $\chi^2$ function gives the confidence level in which tested oscillation parameter values can be ruled out with referenced data. It provides an excellent preliminary evaluation model to estimate the experiment performance.

\begin{table}[h!]
  \centering
 
  \begin{tabular}{ccc}
    \toprule
    Name of Exp & NO$\nu$A & DUNE \\
    \midrule
    Location & Minnesota & South Dakota\\
    POT($yr^{-1}$) & $6.0$x$10^{20}$ & $1.1$x$10^{21}$\\
    Baseline(Far/Near) & 812 km/1km &  1300 km/500 m\\
    Target mass(Far/Near)&14 kt/290 t&40 kt/8 t\\
    Exposure(years)&6&10\\
    Detector type&Tracking Calorimeters&LArTPCs \\
    \bottomrule
  \end{tabular}
    \caption{Details of experiments}
  \label{tab:table1}
\end{table}

\begin{table}[h!]
  \centering

  \begin{tabular}{ccc|c}
\hline
Name Of Exp& Rule & \multicolumn{2}{c}{Normalization error} \\\cline{3-4}

&&signal($\%$)&background($\%$)\\ \hline
&$\nu_{e}$ appearance& 5&10 \\
NO$\nu$A&$\nu_{\mu}$ disappearance& 2&10 \\
&$\overline{\nu}_{e}$ appearance& 5&10 \\
&$\overline{\nu}_{\mu}$ disappearance& 2&10 \\
\hline
&$\nu_{e}$ appearance& 5&10 \\
DUNE&$\nu_{\mu}$ disappearance& 5&10 \\
&$\overline{\nu}_{e}$ appearance& 5&10 \\
&$\overline{\nu}_{\mu}$ disappearance& 5&10 \\
\hline

     \end{tabular}
            \caption{Systematic errors associated with NO$\nu$A and DUNE}
  \label{tab:table2}  
 
\end{table}

\section{Theory}
In a 3+1 sterile neutrino model, the flavour and mass eigenstates are connected through a 4$\times$4 mixing matrix. A convenient parametrization of the mixing matrix is \cite{Adamson:2017zcg}
\begin{equation}
U = R_{34}\tilde{R_{24}}\tilde{R_{14}}R_{23}\tilde{R_{13}}R_{12}.
 \end{equation}
where $R_{ij}$ and $\tilde{R_{ij}}$ represent real and complex 4$\times$4 rotation in the plane containing the 2$\times$2 sub-block in (i,j) sub-block\\\\
\begin{equation}
R_{ij}^{2\times 2}= \left( \begin{array}{cc}
c_{ij} & s_{ij} \\
-s_{ij} & c_{ij} \end{array} \right) \quad\quad 
\tilde{R_{ij}}^{2\times2}= \left( \begin{array}{cc}
c_{ij} & \tilde{s_{ij}} \\
-\tilde{s_{ij}}^* & c_{ij} \end{array} \right) \\\\ \end{equation}
Where, $ c_{ij} = \cos\theta_{ij}$, $ s_{ij} = \sin\theta_{ij}, \tilde{s_{ij}}=s_{ij}e^{-i\delta_{ij}}$ and $\delta_{ij} $ are the CP phases.\\

There are three mass squared difference terms in 3+1 model- $\Delta$m$^2_{21}$(solar)$ \simeq 7.5\times 10^{-5}$eV$^2$, $\Delta$m$^2_{31}$ (atmospheric)$ \simeq 2.4\times 10^{-3}$eV$^2$ and $\Delta$m$^2_{41}$(sterile)$ \simeq 1$eV$^2$. The mass-squared difference term towards which the experiment is sensitive depends on L/E of the experiment. Since SBL experiments have small a very small L/E, $\sin^2(\Delta m^2_{ij}L/4E) \simeq 0$ for $\Delta$m$^2_{21}$ and $\Delta$m$^2_{31}$.  $\Delta$m$^2_{41}$ term survives. Hence, SBL experiments depend only on sterile mixing angles and are insensitive to the CP phases. The oscillation probability, P$_{\mu e}$ for LBL experiments in 3+1 model, after averaging $\Delta$m$^2_{41}$ oscillations and neglecting MSW  effects,\cite{Smirnov:2004zv} is expressed as sum of the four terms
\begin{equation} \label{eq.4}
P_{\mu e} ^{4\nu} \simeq P_1 + P_2(\delta_{13}) + P_3(\delta_{14}-\delta_{24})+P_4(\delta_{13}-(\delta_{14}-\delta_{24})).
\end{equation}

  These terms can be approximately expressed as follows:
\begin{equation} 
\begin{array}{l} \label{eq.5}
\begin{split}
&P_1 = \frac{1}{2}\sin^2 2\theta_{\mu e}^{4\nu}+ [a^2 \sin^2 2\theta_{\mu e}^{3\nu}-\frac{1}{4} \sin^22 \theta_{13}\sin^2 2\theta_{\mu e}^{4\nu} ]\sin^2 \Delta_{31} \\ &+  \big[ a^2b^2 -\frac{1}{4}\sin^22 \theta_{12}(\cos^4\theta_{13} \sin^2 2\theta_{\mu e}^{4\nu}+a^2\sin^2 2\theta_{\mu e}^{3\nu} )]\sin^2 \Delta_{21},
\end{split}
\end{array}
\end{equation}

\begin{equation} 
\begin{array}{l} \label{eq.6}
\begin{split}
&P_2(\delta_{13}) =  a^2b\sin 2\theta_{\mu e}^{3\nu}(\cos2 \theta_{12}\cos\delta _{13}\sin^2 \Delta_{21}-\frac{1}{2}  \sin\delta _{13}\sin 2\Delta_{21}),
\end{split}
\end{array}
\end{equation}

\begin{equation} 
\begin{array}{l} \label{eq.7}
\begin{split}
&P_3(\delta_{14}-\delta_{24}) = ab\sin 2\theta_{\mu e}^{4\nu}\cos^2\theta_{13}\big[\cos2 \theta_{12}\cos(\delta _{14}-\delta _{24}) \sin^2 \Delta_{21} \\ & -\frac{1}{2} \sin(\delta _{14}-\delta _{24})\sin 2\Delta_{21}\big],
\end{split}
\end{array}
\end{equation}

\begin{equation} 
\begin{array}{l} \label{eq.8}
\begin{split}
&P_4(\delta_{13}-(\delta_{14}-\delta_{24})) = a\sin 2\theta_{\mu e}^{3\nu}\sin 2\theta_{\mu e}^{4\nu}\big[ \cos2\theta_{13}\cos(\delta _{13}-(\delta_{14}-\delta_{24})) \sin^2 \Delta_{31} \\ &+\frac{1}{2} \sin(\delta _{13}-(\delta_{14}-\delta_{24}))\sin 2\Delta_{31}  -\frac{1}{4}\sin^22\theta_{12}\cos^2\theta_{13}\cos(\delta _{13}-(\delta_{14}-\delta_{24}))\sin^2 \Delta_{21}  \big],
\end{split}
\end{array}
\end{equation} 
\medskip

With the parameters defined as
\begin{equation} 
\begin{array}{l}  \label{eq.9}
\begin{split}
 \Delta_{ij} &\equiv \Delta m^{2}_{ij}L/4E ,\text{ a function of  baseline(L) and neutrino energy(E)} \\
a &= \cos\theta_{14}\cos\theta_{24}, \\ 
b &= \cos\theta_{13}\cos\theta_{23}\sin2\theta_{12}, \\
\sin 2\theta_{\mu e}^{3\nu} &= \sin2\theta_{13}\sin\theta_{23},\\ 
\sin 2\theta_{\mu e}^{4\nu} &= \sin2\theta_{14}\sin\theta_{24}.
\end{split}
\end{array}
\end{equation}

The CP phases introduced due to sterile neutrinos persist in the P$_{\mu e}$ even after averaging out $\Delta$m$^2_{41}$ lead oscillations. Last two terms of equation \ref{eq.4}, give the sterile CP phase dependence terms. P$_{3}(\delta_{14}-\delta_{24})$ depends on the sterile CP phases $\delta_{14}$ and $\delta_{24}$, while P$_{4}$ depends on a combination of $\delta_{13}$ and $\delta_{14}-\delta_{24}$. Thus, we expect LBL experiments to be sensitive to sterile phases. We note that the probability P$_{\mu e}$ is independent $\theta_{34}$. One can see that $\theta_{34}$ will effect P$_{\mu e}$ if we consider earth mass effects. Since matter effects are relatively small for NO$\nu$A and DUNE, their sensitivity towards $\theta_{34}$ is negligible. The amplitudes of atmospheric-sterile interference term (eq.\ref{eq.8}) and solar-atmospheric interference term(eq.\ref{eq.6}), are of the same order. This new interference term reduces the sensitivity of experiments to the standard CP phase($\delta_{13}$).

\begin{figure}[ht!] 
        \centering \includegraphics[width=1.0\columnwidth]{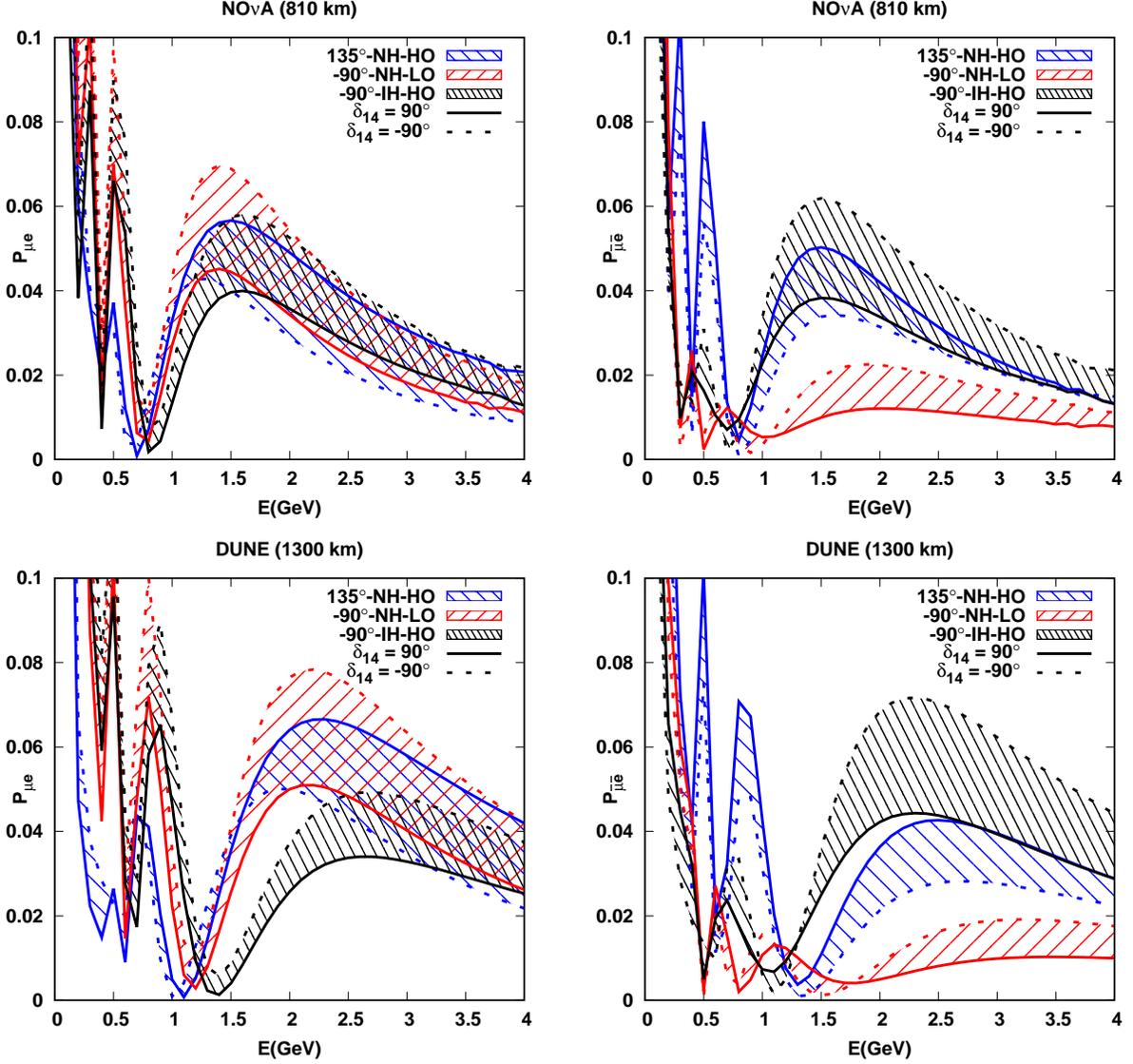}    
        \caption{The oscillation probability P$_{\mu e}$ as a function of energy. The Top(bottom) panel is NO$\nu$A(DUNE). The bands correspond to different values of $\delta_{14}$, ranging from -180$^{\circ}$ to  180$^{\circ}$ when $\delta_{24} = 0^{\circ}$.
Inside each band, the probability for $\delta_{14}$ = 90$^{\circ}$ ($\delta_{14}$ = -90$^{\circ}$) case is shown as the solid (dashed) line. The left(right) panel corresponds to neutrinos(anti-neutrinos).}
\label{fig:1}
\end{figure}

\begin{figure}[ht!] 
        \centering \includegraphics[width=1.0\columnwidth]{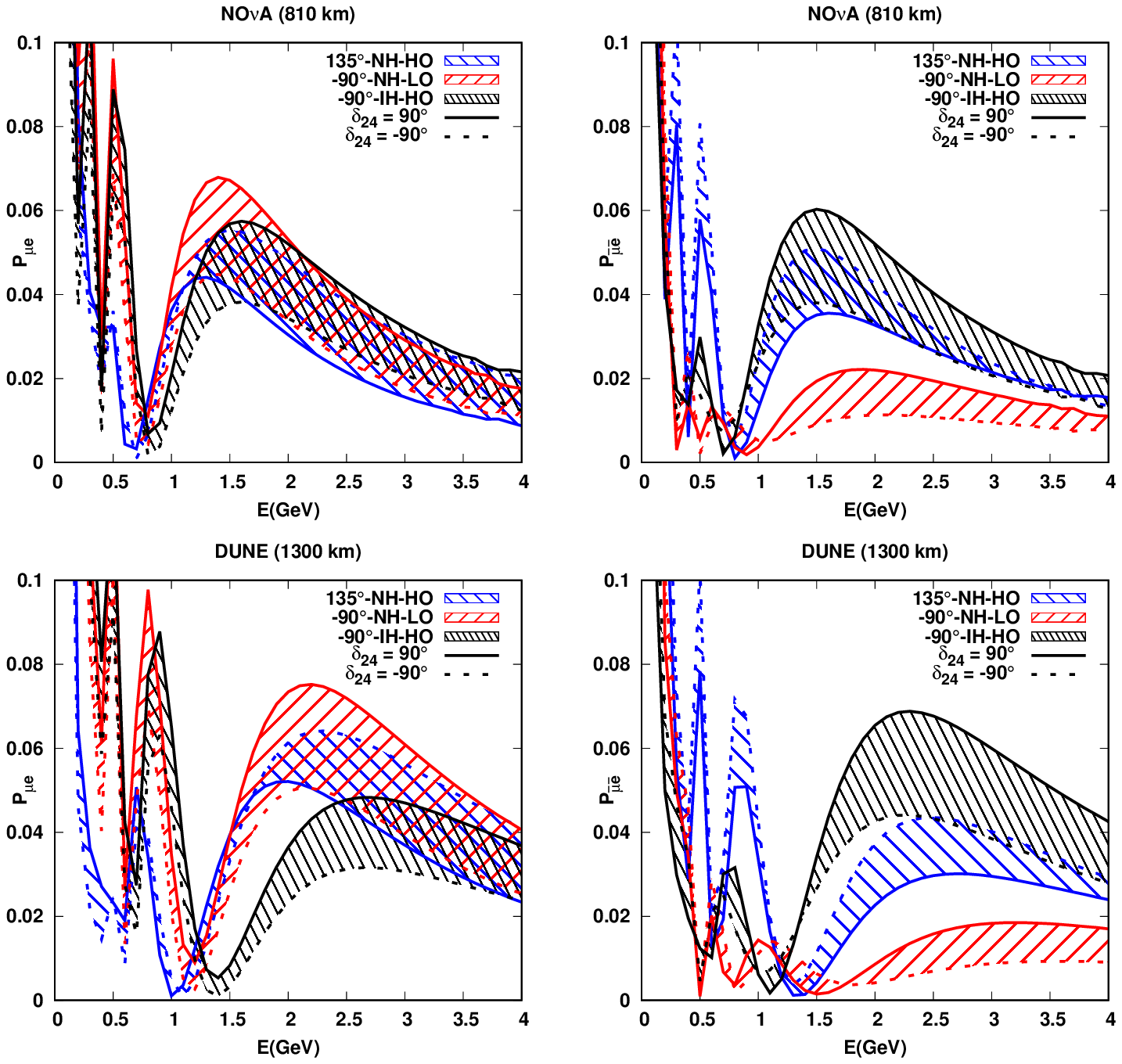}    
        \caption{The oscillation probability P$_{\mu e}$ as a function of energy. The Top(bottom) panel is NO$\nu$A(DUNE). The bands correspond to different values of $\delta_{24}$, ranging from -180$^{\circ}$ to  180$^{\circ}$ when $\delta_{14} = 0^{\circ}$. Inside each band, the probability for $\delta_{24}$ = 90$^{\circ}$ ($\delta_{24}$ = -90$^{\circ}$) case is shown as solid (dashed) line. The left(right) panel is for neutrinos(anti-neutrinos).}
\label{fig:2}
\end{figure}

\begin{table}[h!]
  \centering

  \begin{tabular}{|c|c|c|}
\hline
Parameter& True value & Marginalization Range \\
\hline

sin$^2\theta_{12}$&0.304& Not Marginalized \\
\hline
sin$^2 2\theta_{13}$&0.085& [0.075,0.095] \\
\hline

sin$^2\theta_{23}$&0.623(HO),0.404(LO)& [0.32,0.67] \\
\hline
sin$^2\theta_{14}$&0.025& Not Marginalized \\
\hline

sin$^2\theta_{24}$&0.025& Not Marginalized \\
\hline
sin$^2\theta_{34}$&0.025& Not Marginalized \\
\hline
$\delta_{13}$&135(NH-HO),-90(NH-LO,IH)& [-180,180] \\
\hline
$\delta_{14}$&[-180,180]& [-180,180] \\
\hline
$\delta_{24}$&[-180,180]& [-180,180] \\
\hline
$\Delta m^{2}_{21}$ &$7.50 \times 10^{-5}$ eV$^2$ & Not Marginalized \\
\hline
$\Delta m^{2}_{31}$(NH)&$2.40 \times 10^{-3}$ eV$^2$& Not Marginalized \\
\hline
$\Delta m^{2}_{31}$(IH)&$-2.33 \times 10^{-3}$ eV$^2$& Not Marginalized \\
\hline
$\Delta m^{2}_{41}$ &1 eV$^2$& Not Marginalized \\
\hline

\hline
\end{tabular}
  \caption{Oscillation parameters considered in numerical analysis. The $\sin^2\theta_{23}$ and $\delta_{13}$ are taken from latest NO$\nu$A results\cite{Adamson:2017gxd}.}
    \label{tab:table3}  
\end{table}

\medskip
In figure \ref{fig:1}, we plot the oscillation probability(P$_{\mu e}$) as a function of energy while varying $\delta_{14}$ (-180$^{\circ}$ to 180$^{\circ}$) and keeping $\delta_{24} = 0$ for the three best fit values of latest NO$\nu$A results\cite{Adamson:2017gxd} i.e; NH-LO-1.48$\pi$[$\delta_{13}$], NH-HO-0.74$\pi$ and IH-HO-1.48$\pi$. Where, HO implies $\sin^2\theta_{23} = 0.62$ and LO implies $\sin^2\theta_{23} = 0.40$. For the flux peak of NO$\nu$A, E $\approx$ 2GeV, We observe a degeneracy between all best-fit values due to the presence of $\delta_{14}$ band for neutrino case. While, only NH-HO and IH-HO bands overlap in anti-neutrino case.  We see that $\delta_{14}$ phase decreases both octant and hierarchy resolution capacity for neutrino case and only mass hierarchy resolution capacity for anti-neutrino case. The second row plots P$_{\mu e}$ for DUNE at baseline 1300 km. We observe smaller overlap between bands compared to NO$\nu$A. Thus, the decrease of degeneracy resolution capacity for DUNE is less than NO$\nu$A. Similarly we plot P$_{\mu e}$ while varying $\delta_{24}$(-180$^{\circ}$ to 180$^{\circ}$) in figure \ref{fig:2} and keeping $\delta_{14} = 0^{\circ}$. We see that $\delta_{24}$ has similar effect as of $\delta_{14}$ only change is reversal of $\delta_{24}$ band extrema i.e; $\delta_{24}=-90^{\circ}$ gives same result as $\delta_{14}=90^{\circ}$ and vice versa. This can be explained using equations \ref{eq.4} in which we see $\delta_{14}$ and $\delta_{24}$ are always together with opposite signs. Overall from the probability plots, we observe that the addition of new CP phases decrease octant and mass hierarchy resolution capacities. 

\smallskip
In the next section, we explore how parameter degeneracies are affected in the 3+1 model and the extent to which these degeneracies can be resolved in future runs of NO$\nu$A and DUNE. 
\section{Results for NO$\nu$A and DUNE}
We explore allowed regions in sin$^2 \theta_{23}$-$ \delta_{cp}$ plane from NO$\nu$A and DUNE simulation data with different runtimes, considering latest NO$\nu$A results as true values. Using combined analysis of the disappearance and appearance data, NO$\nu$A reported preferred solutions\cite{Adamson:2017gxd} at normal hierarchy (NH)  with two degenerate best-fit points, one in the lower octant (LO) and $\delta_{cp}$= 1.48$\pi$, the other in higher octant (HO) and $\delta_{cp}$ = 0.74$\pi$. Another solution of inverted hierarchy (IH), 0.46$\sigma$ away from best fit is also reported. Table \ref{tab:table3} shows true values of oscillation parameters and their marginalization ranges we used in our simulation. By studying the allowed regions, we understand the extent to which future runs of NO$\nu$A and DUNE will resolve these degeneracies, if the best fit values are true values.

\begin{figure}[ht!] 
        \centering \includegraphics[width=1.0\columnwidth]{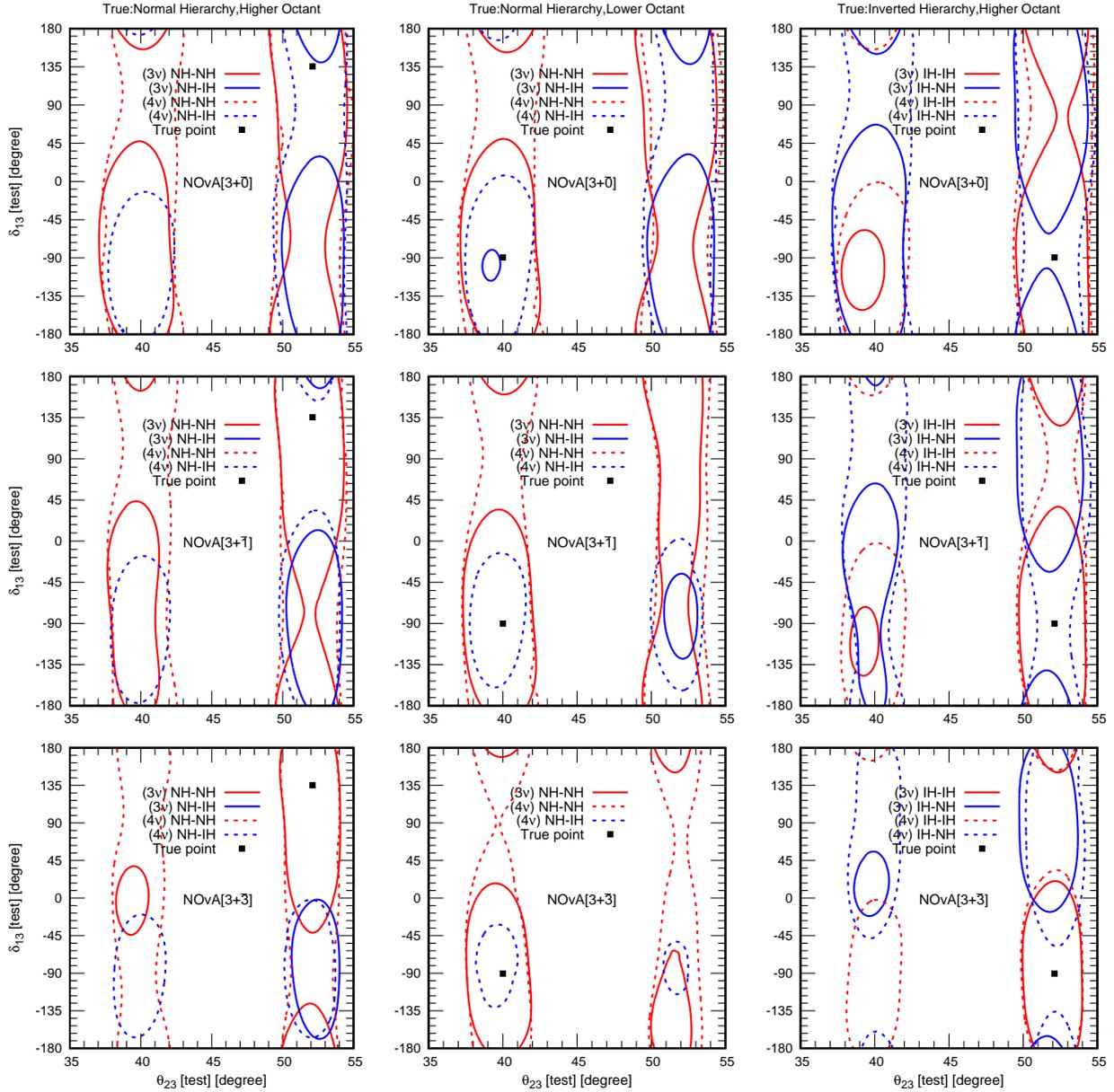}    
        \caption{Contour plots  of allowed regions in the test plane, $\theta_{23}$ vs $\delta_{13}$, at 90$\%$ C.I with top, middle and bottom rows for NO$\nu$A runs of $3+\bar{0},3+\bar{1}$ and $3+\bar{3}$ years respectively.}
\label{fig:3}
\end{figure}

\medskip
In the first row of figure \ref{fig:3}, we show allowed areas for NO$\nu$A[3+$\bar{0}$]. In first plot of first row, we show 90$\%$ C.L allowed regions for true values of $\delta_{13}=135^{\circ}$ and $\theta_{23}= 52^{\circ}$ and normal hierarchy. We plot test values for both NH and IH, of 3 and 3+1 neutrino models. We observe that introducing sterile neutrino largely decreases the precision of $\theta_{23}$. The WO-RH region, for 3$\nu$ case confined between $45^{\circ}$ to $-180^{\circ}$ of $\delta_{13}$, confines the whole $\delta_{13}$ region for 4$\nu$ case. The WH-RO region of 3$\nu$ case doubles, covering the entire region of $\delta_{13}$ for 4$\nu$ case. The 3+1 model also introduces a small WH-WO region, that was absent in 3$\nu$ model.
In the second plot of first row(true value  $\delta_{13}=-90^{\circ}$, $\theta_{23}= 40^{\circ}$ and normal hierarchy), for the 3$\nu$ case, we see RH-RO region excluding $45^{\circ}$ to $150^{\circ}$ of $\delta_{13}$, while RH-WO region covers whole of the $\delta_{13}$ region. In 3+1 model, both RH-RO, RH-WO regions cover whole of the $\delta_{13}$ region. WH-RO solution occupies a small region for 3$\nu$ case, covering half of $\delta_{13}$ region for 4$\nu$ case. WH-WO region covers whole of the $\delta_{13}$ region for 4$\nu$ case. In the third plot of first row, true values are taken as, $\delta_{13}=-90^{\circ}$, $\theta_{23}= 52^{\circ}$ and inverted hierarchy. The RH-RO region covers the entire range of $\delta_{13}$ for both 3$\nu$ and 4$\nu$ case, where as, RH-WO region almost doubles from 3$\nu$ case to 4$\nu$ case. A small range of $\delta_{13}$ excluded from WH-RO for 3$\nu$ case are covered in 4$\nu$ case. WH-WO region of 3$\nu$ case excludes $60^{\circ}$ to $150^{\circ}$ of $\delta_{13}$ while full $\delta_{13}$ range is covered for 4$\nu$ case.

\medskip
In the second row of the figure, we plot allowed regions for NO$\nu$A[3+$\bar{1}$]. We take true values as best fit points obtained by NO$\nu$A. We observe an increase in precision of parameter measurement, due to an increase in statistics, from added 1 yr of anti-neutrino run. In the first plot of the second row, the RH-RO octant region covers entire $\delta_{13}$ range for both 3$\nu$ and 4$\nu$ case. RH-WO region includes $-180^{\circ}$ to $45^{\circ}$ of $\delta_{13}$ for 3$\nu$ case, while whole range of $\delta_{13}$ is covered in 4$\nu$ case. A slight increase in the area of WH-RO is observed form 3$\nu$ to 4$\nu$ case. 4$\nu$ introduces WH-WO region which was resolved for 3$\nu$ case. In the second plot, RH-RO region allows full range of $\delta_{13}$ for 4$\nu$ case, while it was restricted to lower half of CP range in 3$\nu$ case. We see WH-RO solution which was resolved in 3$\nu$ case, is reintroduced in 4$\nu$ case. We also see a slight increase in the size of WH-WO solution from 3$\nu$ to 4$\nu$. In third plot, RH-RO region covers whole CP range for 4$\nu$ while $35^{\circ}$ to $125^{\circ}$ of $\delta_{13}$ are excluded in 3$\nu$ case. The almost resolved RH-WO solution for 3$\nu$ doubles for 4$\nu$ case. WH-RO, WH-WO cover entire region of $\delta_{13}$  for 4$\nu$ case.

\medskip
In the third row, we show allowed regions for NO$\nu$A[3+$\bar{3}$]. In the first plot, it can be seen that small area of RH-WO in case of 3$\nu$ case now covers the whole of $\delta_{13}$ region for 4$\nu$ case. While the 3$\nu$ case has WH-W$\delta_{13}$ degeneracy, 4$\nu$ case introduces equal sized WH-WO-W$\delta_{13}$ degeneracy. In second plot, for 3$\nu$ case: most of $\delta_{13}$ values above $0^{\circ}$ are excluded, but for 4$\nu$ case we see contour covers whole of $\delta_{13}$ range. Already present small area of RH-WO of 3$\nu$ is also increased for 4$\nu$ case. 4$\nu$ case also introduces a small region of WH solutions which were not present in 3$\nu$ case. In the third plot, we see that 4$\nu$ introduces RH-WO region of the almost equal size of RH-RO region of 3$\nu$ case. We observed a slight increase in WH-RO region for 4$\nu$ over 3$\nu$ case, while the WH-WO region almost triples for 4$\nu$ case. 

\begin{figure}[ht!] 
        \centering \includegraphics[width=1.0\columnwidth]{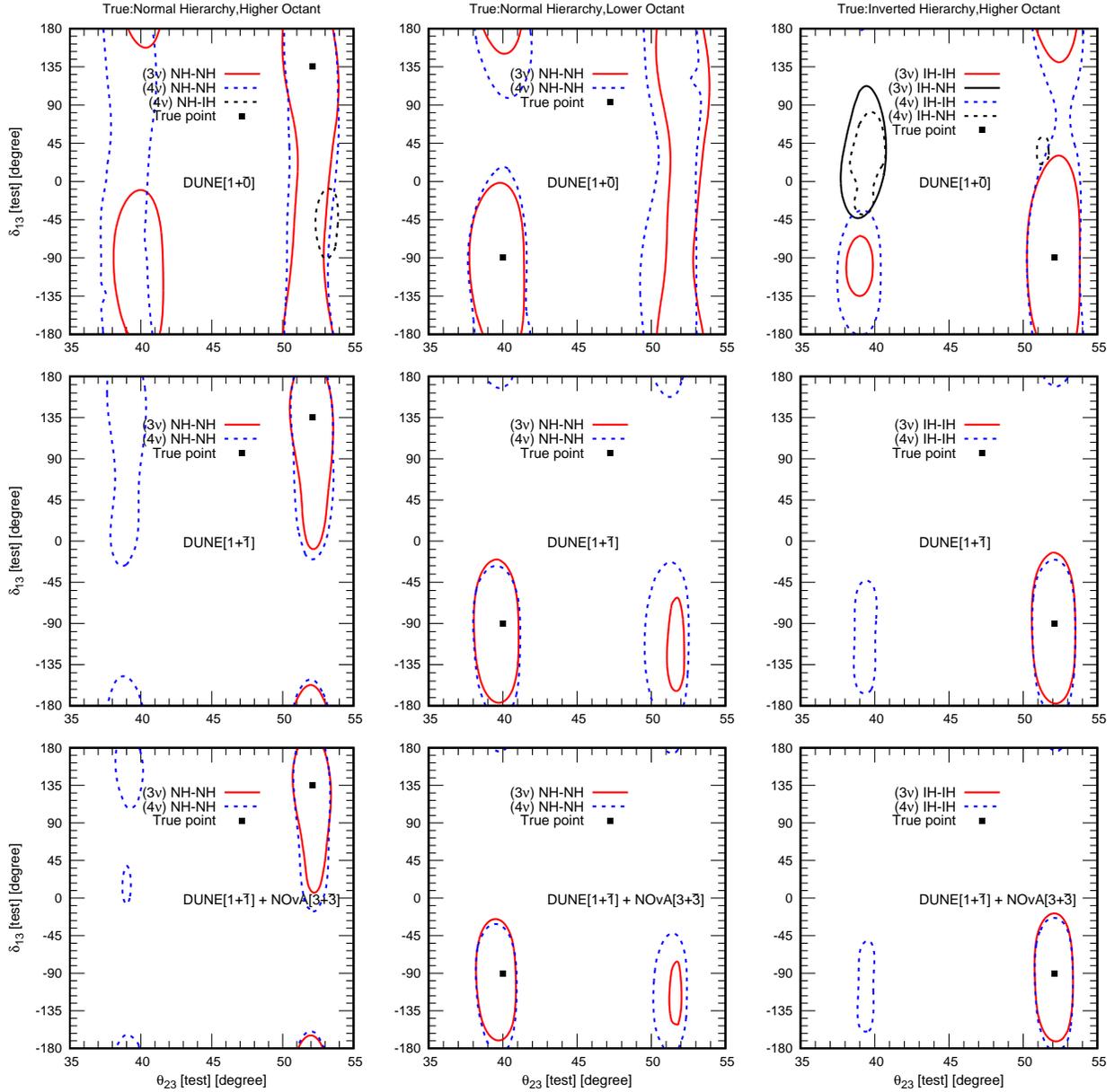}    
        \caption{Contour plots  of allowed regions  in the test plane $\theta_{23}$ vs $\delta_{13}$ at 99$\%$ C.L with top, middle and bottom rows for DUNE runs of $1+\bar{0}, 1+\bar{1}$  years and DUNE[$1+\bar{1}$]+NO$\nu$A[$3+\bar{3}$] respectively.}
\label{fig:4}
\end{figure}

\bigskip
 In the figure \ref{fig:4}, we show allowed parameter regions for DUNE experiment for different run-times. DUNE being the next generation LBL experiment it is expected to have excellent statistics. Hence, We plot 99$\%$ C.L regions for DUNE. In the first row of figure \ref{fig:4}, We show 99$\%$ C.L for DUNE[1+$\bar{0}$]. In the first plot, RH-RO region covers entire of $\delta_{13}$ range for both 3$\nu$ and 4$\nu$ case. The RH-WO region which covers only lower half of $\delta_{13}$ region for 3$\nu$ case covers the whole range for 4$\nu$ case. A small region of WH is also observed. The second plot we see that all WH solutions are resolved. RH-WO covers the whole range of $\delta_{13}$ for both 3$\nu$ and 4$\nu$ case. RH-RO solutions exclude  $0^{\circ}$ to $155^{\circ}$ of $\delta_{13}$ for 3$\nu$ case, while  $20^{\circ}$ to $100^{\circ}$ of $\delta_{13}$ is excluded for 4$\nu$ case. In third plot, we see that 4$\nu$ case extends RH-RO to whole range of $\delta_{13}$ while $30^{\circ}$ to $140^{\circ}$  of $\delta_{13}$ were excluded for 3$\nu$ case. We can see that DUNE clearly has better precision than NO$\nu$A experiment. In the second row, we show allowed regions for DUNE[1+$\bar{1}$]. We see the WH solutions are resolved for both 3$\nu$ and 4$\nu$ cases for all the best-fit values. In the first plot, 4$\nu$ case introduces RH-WO solution of similar size as RH-RO region of 3$\nu$ case. In the second plot, there is no considerable change in 4$\nu$, compared to  3$\nu$ case for RH-RO region, while RH-WO octant is approximately doubled for 4$\nu$ case compared to 3$\nu$ case. In the third plot, 4$\nu$ case introduces small region of RH-WO which covers  $-45^{\circ}$ to $-170^{\circ}$  of $\delta_{13}$. In third row, we combine statistics of  DUNE[1+$\bar{1}$] and NO$\nu$A[3+$\bar{3}$]. There is a small improvement in precision from the combined result over the result from DUNE[1+$\bar{1}$] alone. In the first plot, we see a small RH-WO region is introduced by 4$\nu$ case. In the second plot, there is no considerable change between 3$\nu$ and 4$\nu$ case for RH-RO region, while RH-WO octant almost doubles over 3$\nu$ case for 4$\nu$ case. In the third plot, 4$\nu$ case introduces small region of RH-WO which covers  $-35^{\circ}$ to $-160^{\circ}$  of $\delta_{13}$.

\begin{figure}[ht!] 
        \centering \includegraphics[width=1.0\columnwidth]{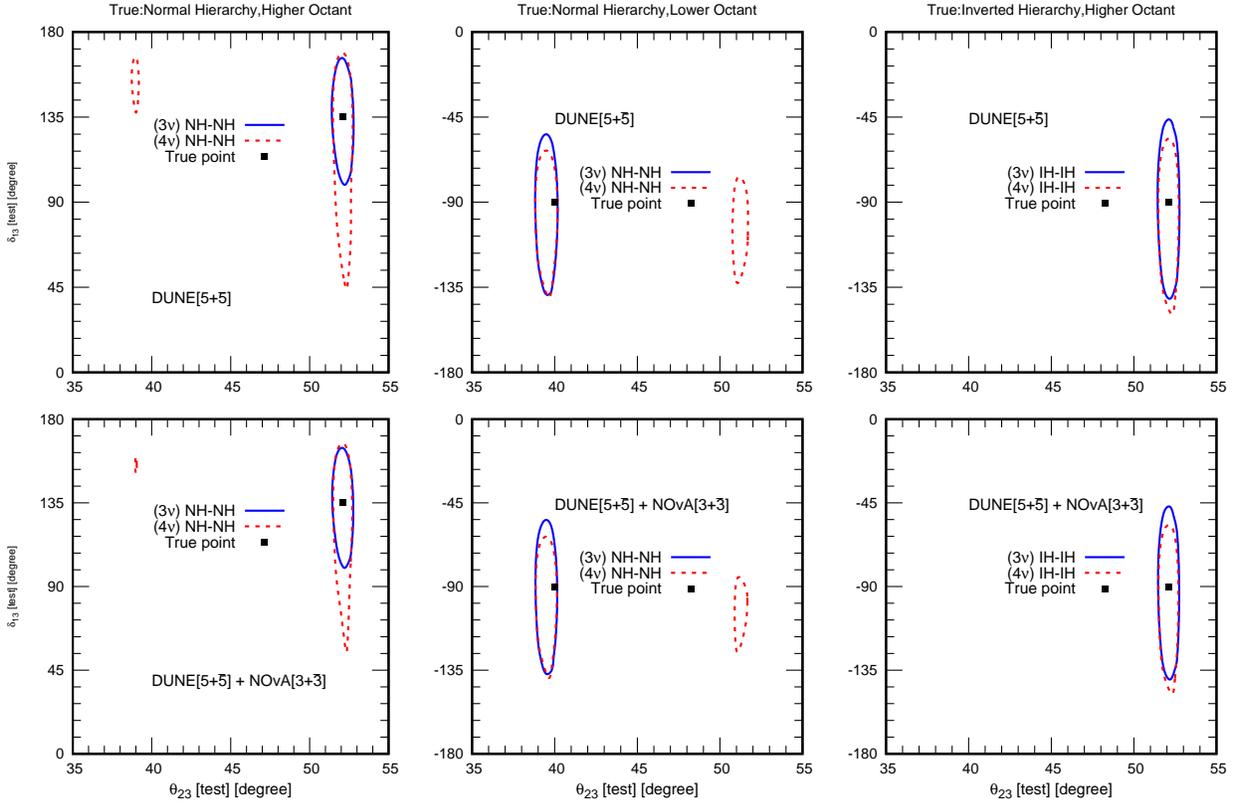}    
        \caption{Contour plots  of allowed regions  in the test plane $\theta_{23}$ vs $\delta_{13}$ at 99$\%$ C.L with top and bottom rows for DUNE[$5+\bar{5}$]and NO$\nu$A[$3+\bar{3}$] + DUNE[$5+\bar{5}$] respectively.}
\label{fig:5}
\end{figure}

In the next figure \ref{fig:5}, we show allowed parameter regions for DUNE experiment, at 99$\%$ C.L for DUNE[5+$\bar{5}$]. We see that WH regions completely disappear for all the true value assumptions. In the first plot, RH-RO region covers a small $\delta_{13}$ range for both 3$\nu$ and 4$\nu$ case indicating high precision measurement capacity of DUNE. We see that $\delta_{13}$ range for 4$\nu$ case is approximately doubled as conpared to the 3$\nu$ case. A small region of RH-WO is observed for 4$\nu$ case. In the second plot, RH-RO region covers small $\delta_{13}$ range of equal area for both 3$\nu$ and 4$\nu$ case. A small region of RH-WO is observed for 4$\nu$ case. In the third plot, the RH-WO solution is resolved. There is an increase in precision due to an increase in statistics. DUNE[5+$\bar{5}$] clearly has a better precision compared to the NO$\nu$A[3+$\bar{3}$] experiment. In the second row, we combine full run of NO$\nu$A and DUNE to check their degeneracy resolution capacity. The WH solutions are resolved for both 3$\nu$ and 4$\nu$ cases for all the best-fit values. In the first plot, RH-WO solution is almost resolved for 4$\nu$ case. In the second plot, RH-RO region covers small $\delta_{13}$ range of equal area for both 3$\nu$ and 4$\nu$ case. A small region of RH-WO is observed for 4$\nu$ case. We observe a slight improvement in degeneracy resolution, on consideration of combined statistics of full run DUNE and NO$\nu$A, over DUNE[5+$\bar{5}$].

\section{Conclusions}
We have discussed how the presence of a sterile neutrino will affect, the physics potential of the proposed experiment DUNE and future runs of NO$\nu$A, in the light of latest NO$\nu$A results\cite{Adamson:2017gxd}. The best-fit parameters reported by NO$\nu$A still contain degenerate solutions. We attempt to see the extent to which these degeneracies could be resolved in future runs for the 3+1 model. Latest NO$\nu$A best-fit values are taken as our true values. First, we show the degeneracy resolution capacity, for future runs of NO$\nu$A. We conclude that NO$\nu$A[3+$\bar{3}$] could resolve WH-WO solutions for first two true value cases, at 90$\%$ C.L for 3$\nu$ case, but not for 4$\nu$ case. DUNE[1+$\bar{1}$] could resolve WH and RH-W$\delta_{cp}$ solutions for both 3$\nu$ and 4$\nu$ case. WO degeneracy is resolved for 3$\nu$ case at 99$\%$ C.L except for small RH-WO region for the second case of true values. DUNE[1+$\bar{1}$] combined with NO$\nu$A[3+$\bar{3}$] shows increased sensitivity towards degeneracy resolution. Finally, for the full planned run of DUNE[5+$\bar{5}$], all the degeneracies are resolved at 99$\%$ C.L for 3$\nu$ case while a tiny region of WO linger on for 4$\nu$ case. For combined statistics of DUNE[5+$\bar{5}$] and NO$\nu$A[3+$\bar{3}$], we observe that all the degeneracies are resolved at 99$\%$ C.L for both 3$\nu$ and 4$\nu$ case except for the NH-LO case. Thus, we conclude that NO$\nu$A and DUNE experiments together can resolve all the degeneracies at 99$\%$ C.L even in the presence of sterile neutrino, if one of the current best-fit values of NO$\nu$A, is the true value.\\\\

\newpage
\textbf{Acknowledgments} 
 AC would like to thank \textit{ The Council of Scientific $\&$ Industrial Research, Government of India}, for financial support. The work of SR was supported by \textit {Department of Science $\&$ Technology, Government of India}. 
 We would like to thank Dr. Monojit Ghosh, Dr. C Soumya and K Siva Prasad for their valuable help.

\end{document}